\documentclass[aip,apl,reprint]{revtex4-1}

\usepackage{graphicx}% Include figure files
\usepackage{dcolumn}% Align table columns on decimal point
\usepackage{bm}% bold math
\usepackage{upgreek} % gives variant characters
\usepackage{siunitx} % typesets values in SI units
\usepackage{xcolor} % for marking up text with colors
\usepackage{textcomp} % gives registered, copyright, etc
\usepackage[utf8]{inputenc} % umlauts
\usepackage{float}% help with precise figure placement
\draft 

\begin{document}
%\preprint{}

\title{Laser-induced electron emission from Au nanowires: a probe for orthogonal polarizations} 

\author{Eric R. Jones}
\email[]{eric.ryan.jones@huskers.unl.edu}
%\thanks{}
\affiliation{Department of Physics and Astronomy, University of Nebraska-Lincoln, Lincoln, Nebraska 68588, USA}

\author{Wayne C. Huang}
% \email[]{wayne.email@address.edu}
%\thanks{}
\affiliation{Department of Physics and Astronomy, University of Nebraska-Lincoln, Lincoln, Nebraska 68588, USA}

\author{Gobind Basnet}
% \email[]{gobind.basnet@phys.ksu.edu}
%\thanks{}
\affiliation{Department of Physics, Kansas State University, Manhattan, Kansas 66506, USA}

\author{Bret N. Flanders}
% \email[]{bret.flanders@phys.ksu.edu}
%\thanks{}
\affiliation{Department of Physics, Kansas State University, Manhattan, Kansas 66506, USA}

\author{Herman Batelaan}
\email[]{hbatelaan@unl.edu}
%\thanks{}
\affiliation{Department of Physics and Astronomy, University of Nebraska-Lincoln, Lincoln, Nebraska 68588, USA}

\date{\today}

\begin{abstract}
Photoelectron field emission, induced by femtosecond laser pulses focused on metallic 
nanotips, provides spatially coherent and temporally short electron pulses. 
Properties of the photoelectron yield give insight into both the material properties 
of the nanostructure and the exciting laser focus.
Ultralong nanoribbons, grown as a single crystal attached to a metallic taper, are 
sources of electron field emission that have not yet been characterized.
In this report, photoemission from gold nanoribbon samples is studied and compared to 
emission from tungsten and gold tips. 
We observe that the emission from sharp tips generally depends on one transverse 
component of the exciting laser field, while the emission of a blunted nanoribbon is 
found to be sensitive to both components. 
We propose that this property makes photoemission from nanoribbons a candidate for 
position-sensitive detection of the longitudinal field component in a tightly 
focused beam.       
\end{abstract}


\maketitle %\maketitle must follow title, authors, abstract and \pacs
%\section{\label{sec:intro} Introduction}
A consequence of tightly focusing a beam of light is that the beam will become longitudinally polarized near the focus.\cite{Richards358,PRA3727,Dorn1917}
% Obtaining longitudinally polarized beams is of particular interest because the focal width can be below the diffraction limit, 
% and the longitudinal component of the electrical field does not contribute to the energy flow along the direction of beam propagation.\cite{Wang501,Yu38859}
Longitudinally polarized beams are desirable because their focal widths can be below 
the diffraction limit, and the longitudinal component of the electric field does not 
contribute to the energy flow along the direction of beam 
propagation.\cite{Wang501,Yu38859} 
These features find applications in high-resolution optical microscopy,\cite{Yoshiki1680,Terakado1114} 
optical data storage,\cite{Kim236} 
particle trapping,\cite{Zhan3377,Zhan6508} 
charged particle acceleration,\cite{Gupta402}
material ablation,\cite{Hnatovsky123901}
and pushing the high-intensity frontier.    
The longitudinal field component of a laser focus has been characterized 
\textit{in situ} by atomic fluorescence\cite{Novotny5251}
and near-field microscopy,\cite{Yu38859} 
and \textit{ex situ} via imaging of material damage\cite{Hnatovsky123901} 
and atomic force microscopy of thin film deformation.\cite{Gilbert613} 
As the \textit{in situ} methods of characterizing the longitudinal field are limited 
by intensity or to a resonant wavelength, a flexible alternative would be preferable. 

Photoelectron field emission, induced by focusing femtosecond laser pulses onto sharp 
metallic tapers with nanometric radii of curvature,
\cite{Hommelhoff077401,Ropers043907,Barwick142}
has a broad range of applications. 
% For moderate intensities, it is possible to produce temporally short electron wave 
% packets\cite{Kruger78} 
% with high spatial coherence.\cite{Hommelhoff423,Ehberger227601} 
Temporally short electron wave packets\cite{Kruger78} with high spatial coherence\cite{Hommelhoff423,Ehberger227601} 
can be achieved with moderate intensities. 
Tip sources have thus been integrated into electron microscopes to obtain 
% timing on the order of hundreds of femtoseconds and 
sub-micron spatial resolution with femtosecond timing.\cite{Barwick902,Gulde200,Piazza6407} 
Femtosecond electron pulses have been used to study fundamental quantum mechanics, as 
in testing the existence of forces in the Aharonov-Bohm effect,\cite{Caprez210401}
and observing diffraction in time.\cite{Kruger78,Kruger074006}
Electron emission from nanotips is obtained for a range of laser intensities and 
wavelengths.\cite{Herink123005,Forster217601,Huang023011} 
Nanotip emission might then provide an alternative for characterizing the longitudinal 
component of a laser focus.  
However, as nanotip emission is dominated by a single transverse component of the 
polarization of the exciting field, it has not yet been utilized as a detector 
of longitudinal fields. 
 
Ultralong gold nanoribbons,\cite{Basnet073106}
grown by the method of directed electrochemical nanowire assembly 
(DENA),\cite{Ozturk175707,Flanders1130001}are as of yet unstudied sources for ultrafast
electron photoemission. 
The DENA methodology results in single-crystalline samples, as confirmed by electron 
diffraction.  
Previous studies into the optical damage threshold of similar nanowire samples 
indicated that single-crystalline gold nanowires could tolerate high laser pulse peak 
intensities before melting, but that they exhibited long cooling 
times.\cite{Summers4235} 
From these studies, the exchange and dissipation of heat between the nanowire electrons
and the crystalline lattice could be modelled. 
A logical next step would then be to characterize electrons emitted from nanoribbons 
in response to ultrashort pulse illumination. 
Photoelectron emission is known to carry information on material effects, such as 
plasmonic dynamics and laser heating.\cite{Vogelsang4685,Kealhofer035045} 
Temporally short electron pulses have been observed from plasmonic 
nanostructures,\cite{Irvine184801,Dombi24206} and the photoemission yield has been used
as a sensitive probe of the plasmonic field enhancement from 
nanostructures.\cite{Racz1181}     
We measured photoelectron emission from nanoribbons in an attempt to determine their 
plasmonic or thermal properties.\cite{Vogelsang4685,Racz1181,Kealhofer035045} 
In the following, nanoribbon samples are characterized by photoelectron emission and 
compared to standard single-crystalline gold and tungsten samples. 
The resulting photoelectron emission spectra reveal that nanoribbons can be employed as 
position- and polarization-sensitive detectors within a laser focus,
providing a potential \textit{in situ} sub-wavelength probe for longitudinal 
polarization.

\begin{figure}
\includegraphics{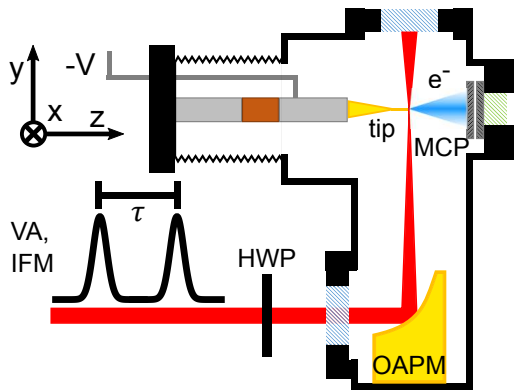}
\caption{\label{fig:fig1} Experimental schematic showing pump/probe configuration for tip and nanoribbon photoelectron emission.
%Element labels are described below in the text.       
}
\end{figure}

A schematic for the system used to characterize tip samples is given in Fig.~\ref{fig:fig1}.
The intensity of the output from a Ti:Saph oscillator 
(Spectra Physics Tsunami, \SI{80}{\MHz} repetition rate, \SI{800}{\nm} central 
wavelength, \SI{100}{\fs} pulse width) is controlled by a variable attenuator (VA), 
which consists of a $\lambda/2$ plate and a Brewster window. 
The attenuated beam is split into a pump and a probe beam by a balanced Mach-Zehnder 
interferometer (IFM).
Photoelectron emission is observed in pump/probe and single beam experiments.  
The time delay between pump and probe pulses, $\uptau$, can be varied from \SI{-4}{\ps}
to \SI{4}{\ps} manually by a micrometer.
The additive ratio, measured as a function of the delay $\uptau$, is obtained by 
dividing the emission rate measured with both pump and probe 
pulses delivered to the samples by the sum of the individual rates due to the pump and 
probe pulses. 
A second $\lambda/2$ plate (HWP) rotates the polarization of both beams prior to 
delivery to the experimental chamber. 
A rotational stepper motor is used to scan the beam power and the polarization.

The experimental chamber, which is detailed in Ref.~\onlinecite{Barwick142}, is 
maintained at \num{2d-7} Torr.   
The beams are focused within the chamber by an off-axis parabolic mirror (OAPM) to a 
full-width half maximum of \SI{3.6}{\um}. 
A 3-axis stage, coupled to the chamber by flexible bellows, positions tip samples into 
the focus.
Mounted tip samples were biased at \SI{-100}{\V}, as this was lower than the threshold 
for Fowler-Nordheim field emission. 
Electrons were collimated through two \SI{4}{\mm} apertures before being detected by a 
microchannel plate (MCP). 

Electron pulses from the MCP were amplified and discriminated. 
Discriminator pulses were counted by a multichannel scaler, and used as the start 
trigger for a time-to-amplitude converter (TAC). 
The output reference signal from the Ti:Saph oscillator was used as the TAC stop 
trigger to measure the arrival time of electrons.
Timing spectra were obtained by sending the TAC output pulses to a multichannel 
analyzer (MCA). 

Nanoribbon samples were prepared using the DENA methodology.
\cite{Ozturk175707,Flanders1130001}
Nanoribbon samples are reported to have a thickness of \SI{40}{\nm}, and widths ranging 
from \SI{130}{\nm}--\SI{360}{\nm} along the length of the wire. 
The tip can have a radius of curvature of \SI{10}{\nm}.\cite{Basnet073106}
These dimensions can be tailored during the growth process to make nanoribbons that are
well-suited for photoemission.
 
In order to distinguish which photoemission properties arise due to the material or 
geometry of the nanoribbon samples, single-crystal tungsten (W) and gold (Au) tips were 
prepared for comparison. 
Samples of W wire (\SI{200}{\um} diameter) were annealed\cite{Greiner026104}, 
and then etched via the lamella drop-off method.\cite{Mueller3970} 
Au wire samples (\num{99.95}\% purity, Ted Pella, \SI{200}{\um} diameter) were annealed\cite{Roy631},
and then etched as according to Refs.~\onlinecite{Gingery113703,Eligal033701}. 

\begin{figure*}
\includegraphics{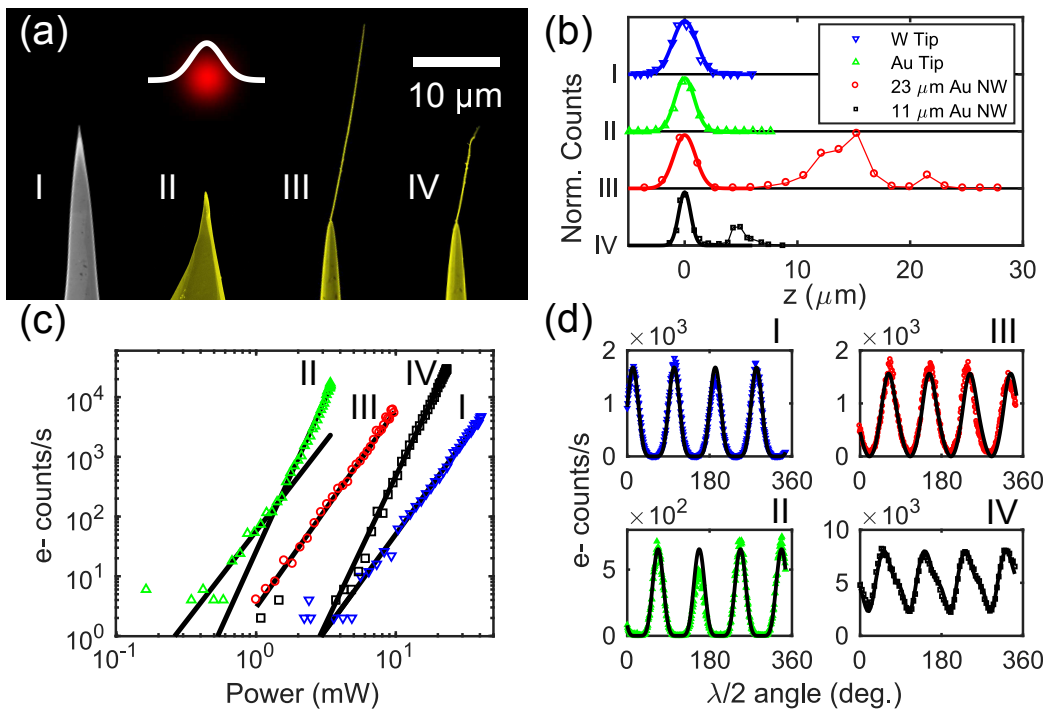}
\caption{\label{fig:fig2} Tip sample characteristics.
(a) Scanning electron microscope (SEM) images of the tip samples:
from left to right is shown annealed W (I), annealed Au (II), 
a \SI{23}{\um} Au nanoribbon (III), and an \SI{11}{\um} Au nanoribbon (IV). 
The laser spot size (red) and intensity profile (white) have a full width half maximum 
of \SI{3.6}{\um} as fit from the W data in (b) and (c).  
(b) Scaled electron counts as function of tip position in focus. Gaussian fits to the data are shown in bold lines. Thin lines are guides to the eye.  
(c) Power dependence of electron emission.  
(d) Polarization dependence of electron emission.
}
\end{figure*}

The tip and nanoribbon samples were mounted to SEM pin stubs with silver paste. 
SEM images of the samples are given in Fig.~\ref{fig:fig2}(a) along with plots of the beam focus (red) and intensity profile (white) as measured by photoemission. 
From left to right is shown W (I), Au (II), an undamaged \SI{23}{\um} Au nanoribbon (III), and an \SI{11}{\um} Au nanoribbon (IV) obtained after the 
\SI{23}{\um} nanoribbon was blunted during pump/probe experiments. 
Images were taken before and after experimental characterization to determine the extent of damage due to laser illumination.  
  
Photoemission data from single-beam experiments are shown in Fig.~\ref{fig:fig2}(b)--(d).
In Fig.~\ref{fig:fig2}(b), the emission rate is shown as samples were translated through the laser focus.
The W (I) and Au (II) tip samples show emission localized at the tip apex only, while the nanoribbon samples (III and IV) can emit from multiple locations along their length.
This feature confirmed that a nanoribbon remained attached to the Au substrate after imaging and transfer to the experimental chamber.
Thin lines between data points serve as a guide to the eye.  
Fig.~\ref{fig:fig2}(c) shows the dependence of electron yield on the average power of the beam, plotted on a log-log scale. 
The value, $n$, of the power dependence ($\propto I^{n}$), is often used to identify the emission process of a tip. 
The W tip and \SI{23}{\um} nanoribbon have slopes of $n=3$, while the Au tip is found 
to have a slope of $n=3$ for low power, and $n=5$ for higher power.
Such behavior, that is, the increase in power law slope with increasing laser power, 
has been observed in W tips and studies of above threshold 
photoemission.\cite{Barwick142,Schenk257601}     
The \SI{11}{\um} nanoribbon has a slope of $n=5$.
With these values of $n$, the position dependence of the samples in Fig.~\ref{fig:fig2}
(b) can be fit with a gaussian function to determine the size of the focal waist (bold lines). 
% A fit of the W position dependence with $n=3$ gives a full width at half maximum of \SI{3.6}{\um} for the focal waist.      
The focal waist has a fitted full width at half maximum of \SI{3.6}{\um} from the W data.
Fig.~\ref{fig:fig2}(d) shows the variation of emission rate as the polarization of the beam is rotated by a $\lambda/2$ plate. 
The high contrast \ang{90} spaced peaks in the tip samples and \SI{23}{\um} nanoribbon support that the sample geometry is well-defined with respect to the 
laser polarization in the focus, and that the emission process is dominated by a preferred laser polarization. 
The broadened 
peaks and reduced contrast of the \SI{11}{\um} nanoribbon electron yield (IV) indicate that the emission process depends on both transverse components of 
the exciting field.
A feature consistent with multiphoton emission is that the power law slope, $n$, will agree with the polarization dependence on the emission rate.
The emission rate is $\propto cos^{2n}(\theta)$ in a multiphoton emission model,
where $\theta$ is the polarization of the laser relative to the tip direction.
The W and Au polarization data agree with a fit with $n=3$, while both nanoribbon samples require a combination of $n=1$ and $n=3$.
This peculiarity of both nanoribbon samples indicates a deviation from typical multiphoton emission.   

\begin{figure*}
\includegraphics{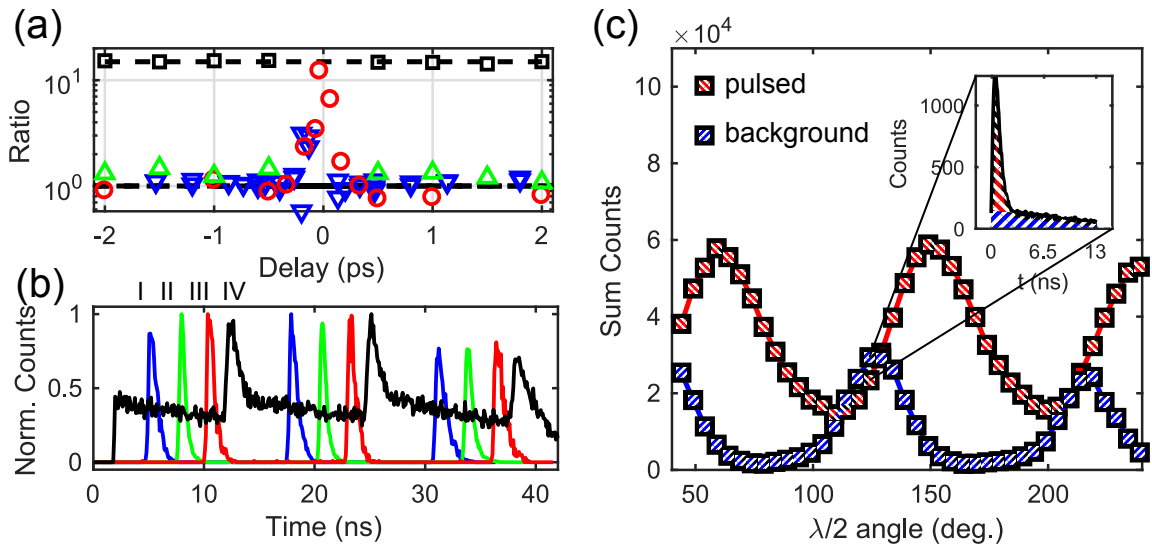}
\caption{\label{fig:fig3} 
Pump/probe, timing, and polarization control of emission processes.
(a) Additive ratio of electron emission vs. pulse delay. 
Color and marker conventions follow Fig.~\ref{fig:fig2}.
Ratio values of 1 and 14.9 are highlighted by dashed black lines. 
(b) Normalized time spectra of electron emission from tip samples. 
Grouped from left to right are the W tip (blue), Au tip (green), \SI{23}{\um} nanoribbon (red), and \SI{11}{\um} nanoribbon (black).
% The peak separation for an individual sample shows the 13 ns repetition rate of the Ti:Saph oscillator.    
(c) Polarization dependence of summed counts of time spectra background (blue hatched) and pulsed (red hatching) emission vs. $\lambda/2$ angle for the \SI{11}{\um} nanoribbon. 
Colored lines are guides to the eye. 
The inset shows the background and pulsed signal for a single timing peak taken at $\lambda/2$ angle $= 124^{\circ}$.
}
\end{figure*}

Pump/probe and single beam experiments with the TAC/MCA configuration revealed the timing features of electron emission. 
The additive ratio of emission from the samples is plotted in Fig.~\ref{fig:fig3}(a) as a function of $\uptau$.
The polarization of the focus was chosen for the optimum electron yield from each sample. 
A ratio of 1 indicates that the emission yield from the probe pulses are independent from the pump pulses.
A ratio significantly greater than 1 indicates emission processes that are slower than the time delay between pulses.\cite{Barwick142} 
When the delay is shorter than the pulse duration, the additive ratio can vary due to interference between the pulses. 
The W tip (blue triangles), Au tip (green triangles), and \SI{23}{\um} nanoribbon all have additive ratios that are close to 1 when the pulse delay is  
outside of the $\pm \SI{200}{\fs}$ interference window, so the emission processes are as fast as the ~\SI{100}{fs} laser pulse duration and thus prompt.
The \SI{11}{\um} nanoribbon (black squares) has an average additive ratio of 14.9 for delays longer than the pump/probe interference window, therefore the process is not prompt. 
Measurement of the ratio for delays with high constructive interference were avoided 
to prevent damage to the Au tip and \SI{11}{\um} nanoribbon samples.      

Normalized time spectra of electron emission from tip samples are plotted in Fig.~\ref{fig:fig3}(b). 
Shown, grouped from left to right, are the spectra from the W tip (blue line), Au tip (green line), \SI{23}{\um} nanoribbon (red line), 
and the \SI{11}{\um} nanoribbon (black line). 
The peak separation for each sample shows the \SI{13}{\ns} pulse separation of the oscillator. 
Sharp peaks indicate \textit{pulsed} electron emission, while sustained signal after the laser pulse indicates \textit{background} emission.  
The \SI{11}{\um} nanoribbon has a significant background as compared to the other samples, indicating that electron emission is continuing after the exciting laser pulse is gone.
The emission process of the \SI{11}{\um} nanoribbon is therefore ruled out as purely multiphoton, and is likely due to both multiphoton and laser heating of the nanoribbon structure.  

To further investigate this feature, the pulsed and background contributions to 
emission from the \SI{11}{\um} nanoribbon are plotted as a function of $\lambda/2$ 
angle in Fig.~\ref{fig:fig3}(c).
Time spectra were recorded for each $\lambda/2$ angle. 
The process for dividing each spectrum into pulsed and background contributions is 
illustrated in the inset of Fig.~\ref{fig:fig3}(c), 
which shows a portion of the time spectrum taken at $\lambda/2$ angle of \ang{124}. 
The red hatched region of the inset indicates the pulsed contribution, and the blue 
hatched region indicates the background.
The background regions are defined by taking linear fits to the tails of the timing 
spectra, and extending those fits to the rising edges of the timing peaks.
This procedure is performed for two \SI{13}{\ns} oscillator periods.
The counts in the background regions are summed, giving the data points marked by the 
blue hatched squares.
The background contributions are then subtracted from the total counts in each 
spectrum. 
This results in the data points marked by the red hatched squares. 
The maximum emission for the background process occurs at a $\lambda/2$ angle that is 
shifted relative to the pulsed process.
For example, for a $\lambda/2$ angle of \ang{79}, the photoemission signal is dominated
by the pulsed process, as the number of counts in the tails of the electron time of 
flight spectra is low.
For a $\lambda/2$ angle of \ang{124}, the photoemission signal is dominated by the 
background process, which is marked by a comparatively high number of counts in the 
delayed tail of the electron time of flight spectrum.
The explanation for the variation in electron signal is likely due to the nanoribbon 
absorbing more of the incident laser pulse energy at \ang{124}, and less at \ang{79}.
The higher degree of energy absorption in the \SI{11}{\um} nanoribbon requires a higher
degree of energy dissipation, which occurs by an additional process--thermal 
dissipation--that is much slower than multiphoton-driven processes. 
This interpretation is consistent with the results of 
Ref.~\onlinecite{Kealhofer035045}, which demonstrated polarization control of thermally 
enhanced photoemission from nanotips. 
The polarization control of these multiphoton (pulsed) and thermal (background) 
processes indicates that they respond to different components of the laser field in the
focus. 

\begin{figure}
\includegraphics{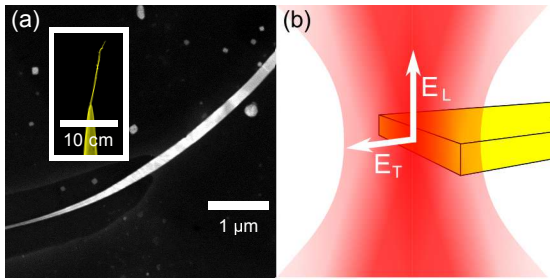}
\caption{\label{fig:fig4} 
(a) Micrographs of a long nanoribbon and sample (IV) (inset).
The nanoribbon is twisted to highlight that it is ribbon shaped. 
(b) Schematic of nanoribbon as a detector of longitudinal ($E_L$) and transverse 
($E_T$) electric fields. 
}
\end{figure}

Ultralong Au nanoribbons are unique nanostructures for the study of electron, thermal, 
and plasmonic transport by laser-induced electron emission. 
Previous work indicated that nanoribbons are resilient to damage by laser intensities 
on the order of \SI[per-mode=symbol]{}{\tera\watt\per\centi\metre\squared},\cite{Summers4235}
but we found that damage can occur with lower intensities. 
This observation resulted in the \SI{11}{\um} nanoribbon, which differed from the 
original \SI{23}{\um} nanoribbon by having a less defined apex.
An immediate consequence of the laser damage to the \SI{23}{\um} nanoribbon was that
the shorter \SI{11}{\um} nanoribbon required higher incident laser power to emit, which
is evident in Fig.~\ref{fig:fig2}(c). 
The change in the nanoribbon's apex geometry also coincided with emission from the 
\SI{11}{\um} nanoribbon being superadditive for emission rates similar to the other
samples. 
The superadditive emission in pump/probe experiments was accompanied by delayed tails 
in the photoelectron time of flight spectra in the \SI{11}{\um} nanoribbon.
Superadditive and delayed emission are not consistent with plasmon-induced field 
emission, as plasmonic emission is reported to have a standard pump/probe 
cross-correlation and thus is as fast as the exciting laser 
pulses.\cite{Dombi24206,Vogelsang4685}   
Such tails were not observed at the \SI{23}{\um} nanoribbon apex before it was damaged.
These features, the superadditive emission and the delayed arrival times, are 
consistent with the \SI{11}{\um} nanoribbon being more susceptible to laser 
heating than the \SI{23}{\um} nanoribbon. 
The higher susceptibility to heating made the \SI{11}{\um} nanoribbon sensitive to both
transverse components of the focused laser field, as shown in Fig.~\ref{fig:fig3}(c).
The sensitivity to both transverse components of the focused field is a unique feature 
of the thermal emission of the nanoribbon that was not observed in W or Au tips.
The cone structure of the nanotip samples leads to much faster cooling times than the 
nanoribbon samples.
This suggests that a tailored nanoribbon could be oriented to probe other polarization
components in a focus as well.   
A schematic for an oriented nanoribbon as a probe of the longitudinal component of a focused non-paraxial beam is given in Fig.~\ref{fig:fig4}.
Shown in Fig.~\ref{fig:fig4}(a) is a high-resolution SEM image of a nanoribbon and our 
\SI{11}{\um} nanoribbon as the inset.       
The nanoribbon is twisted in the high-resolution image to highlight its shape. 
With the nanoribbon oriented as shown in Fig.~\ref{fig:fig4}(b), the broad side of the 
nanoribbon would heat depending on the strength of the longitudinal component of the 
exciting field, $E_L$, and the delayed electron yield would therefore depend on $E_L$,
while the peaked electron yield could depend on a combination of the transverse 
component, $E_T$, and also $E_L$. 
The capability to distinguish the slow thermal electron yield at the nanoribbon apex 
makes electron emission from a nanoribbon a subwavelength probe of orthogonal 
polarizations.     

\begin{acknowledgments}
We gratefully acknowledge funding by NSF EPSCoR NE-KS Track-II, Award No. EPS 1430519, and NSF grant No. 1602755.
\end{acknowledgments}

\bibliography{aipjones_nanowire}

\end{document}